**Kinetic-Scale Topological Structures Associated with Energy Dissipation in the Turbulent Reconnection Outflow**

S. Y. Huang[1][*], J. Zhang[1], Q. Y. Xiong[1], Z. G. Yuan[1], K. Jiang[1], S. B. Xu[1], Y. Y. Wei[1], R. T. Lin[1], L. Yu[1], and Z. Wang[1]

[1]School of Electronic Information, Hubei Luojia Laboratory, Wuhan University, Wuhan, 430072, China

[*]Corresponding author: S. Y. Huang (shiyonghuang@whu.edu.cn)

**Abstract**

Assisted with Magnetospheric Multiscale (MMS) mission capturing unprecedented high-resolution data in the terrestrial magnetotail, we apply a local streamline-topology classification methodology to investigate the categorization of the magnetic-field topological structures at kinetic scales in the turbulent reconnection outflow. It is found that strong correlations between the straining and rotational part of the velocity gradient tensor as well as the magnetic-field gradient tensor. The strong energy dissipation prefers to occur at regions with high magnetic stress or current density, which is contributed mainly by O-type topologies. These results indicate that the kinetic structures with O-type topology play more import role in energy dissipation in turbulent reconnection outflow.

**1. Introduction**

Magnetic reconnection and turbulence are both fundamental processes in space, astrophysical and experimental plasmas. Magnetic reconnection is one of the most crucial energy conversion processes in which magnetic energy is converted to heat and accelerate the particles (e.g., Deng & Matsumoto 2001; Burch et al., 2016a; Huang et al., 2012, 2018; Fu et al., 2016, 2017; Jiang et al., 2019, 2022; Torbert et al., 2018). Turbulence is a ubiquitous nonlinear phenomenon where chaotic dynamics and power-law statistics coexist (e.g., Tu & Marsch 1995; Bruno & Carbone 2013; Sahraoui et al., 2009, 2020; He et al., 2011, 2015; Huang et al., 2014, 2021; Huang & Sahraoui, 2019). Magnetic reconnection is intertwined with turbulence across scales. Specifically, turbulence can dissipate magnetic energy through magnetic reconnection at small-scale thin current sheets (e.g., Retinò et al., 2007; Sundkvist et al., 2007; Phan et al., 2018). In turn, large-scale magnetic reconnection can trigger turbulence generation in the outflows (e.g., Eastwood et al., 2009; Daughton et al., 2011; Huang et al., 2012, 2022; Zhou et al., 2017, 2021; Ergun et al., 2018). Studying the turbulence properties developed in the outflows of magnetic reconnection can further fulfill the understanding of the connection between turbulence and reconnection (e.g., Lapenta et al., 2015, 2016).

From the energy perspective, turbulence is a practical approach to complete the cascade process and bring the energy from large scale to small scale. One hot topic of turbulence studies is unraveling the structures associated with energy dissipation and its mechanisms. Many coherent structures have been stated to be the potential sites where turbulent dissipation occurs, such as thin current sheets, electron vortex magnetic holes, magnetic islands or flux ropes (e.g., Bruno & Carbone 2013; Huang et al., 2016, 2017). Most recently, the tight relationship between intermittent currents and Ohmic dissipation or heating (e.g., Osman et al., 2010; Wan et al., 2012; Chasapis et al., 2015, 2018) has been demonstrated in the space plasmas studies. Besides, a similar correlation between kinetic dissipation expressed by incompressible pressure-strain interactions $\mathbf{\Pi} \cdot \mathbf{D} = (\mathbf{P} - p\mathbf{I}) \cdot (\frac{1}{2}\nabla \mathbf{V} + \frac{1}{2}\nabla \mathbf{V}^{\mathrm{T}} - \frac{1}{3}\nabla \cdot \mathbf{V}\mathbf{I})$ (here $\mathbf{P}$ is the pressure tensor, $p = \frac{1}{3}tr(\mathbf{P})$ is the scalar pressure, $\mathbf{I}$ is the unity tensor, $\mathbf{V}$ is the velocity) and enhanced vorticity or symmetric stress has also been found (e.g., Yang et al., 2017; Bandyopadhyay et al., 2020). These kinetic-scale structures can signal the energy dissipation as those close connections. Moreover, these kinds of structures may be responsible for the energy dissipation in the turbulent reconnection outflow (e.g., Huang et al., 2022; Li et al., 2022).

The energy dissipation is highly correlated with the structure categorizations in the turbulent outflow. It is found that intermittent dissipation by turbulence reconnection and especially energy dissipation in magnetic reconnection occurs at O-lines but not X-lines (e.g., Fu et al., 2017). Also, secondary reconnections can occur in the turbulent outflow and dissipate the magnetic energy therein (e.g., Zhou et al., 2021). Furthermore, the intermittent dissipation at the kinetic scale occurs in the turbulent reconnection outflow, where strong energy dissipation occurs in the intermittent structures and the region with strong current filaments (e.g., Huang et al., 2022; Li et al., 2022).

Recently, a methodology constructed by vector-field gradient tensor could reveal structures and dynamics in turbulence using a series of geometrical invariants (e.g., Chong et al., 1990; Martín et al., 1998; Meneveau, 2011). It has been applied in local streamline-topology classification in both magnetohydrodynamic (MHD) turbulence simulations (e.g., Dallas & Alexakis, 2013) and *in-situ* observations in the turbulent solar wind (e.g., Quattrociocchi et al., 2019; Hnat et al., 2021), turbulent Earth's magnetosheath (e.g., Consolini et al., 2015; Zhang et al., 2023; Ji et al., 2023) and magnetic reconnection region (e.g., Consolini et al., 2018). In order to advance the understanding of turbulent reconnection outflows in the terrestrial magnetotail, the aim of this study is to apply the topology-classification methodology to in-situ observations from Magnetospheric Multiscale (MMS) mission and reveal the most relevant local field topology associated with energy dissipation. This paper is organized as follows: In section 2, we provide a concise yet comprehensive introduction to the concept of gradient tensor and geometrical invariants, which serve as fundamental tools for characterizing and analyzing intricate magnetic field structures. Building upon this foundation, in section 3, we present the data acquired from observations and unveil the corresponding results obtained through the application of our methodology. Finally, in section 4, we summarize and engage in an insightful discussion of the findings.

## 2. Brief introduction of gradient tensor and geometrical invariants

Considering the magnetic field gradient tensor $\mathbf{X} = \nabla \mathbf{B}$, the tensor principal invariants are independent of the frame of reference and are the coefficients of the characteristic polynomial

$$|\mathbf{X} - \lambda_i \mathbf{I}| = \lambda_i^3 + P_X \lambda_i^2 + Q_X \lambda_i + R_X = 0, \qquad (1)$$

where $\lambda_i$ are the eigenvalues of $X$. Three geometric invariants are expressed as

$$P_X = -\text{tr}(\mathbf{X}), \qquad (2)$$

$$Q_X = \frac{1}{2}[P_X^2 - \text{tr}(\mathbf{X}^2)], \qquad (3)$$

$$R_X = -\frac{1}{3}\text{tr}(\mathbf{X}^3). \qquad (4)$$

The discriminant of the characteristic equation for any traceless second-order tensor ($\text{tr}(\mathbf{X}) \equiv \nabla \cdot \mathbf{B} = 0$) in analogy with $\mathbf{X}$ is

$$D_X = \frac{27}{4} R_X^2 + Q_X^3, \qquad (5)$$

providing a general classification for the solutions of Eq. (1) as well as defining two regions in the $(R_X, Q_X)$ plane. When $D_X > 0$, there are two complex-conjugate eigenvalues and one real eigenvalue, and thus magnetic field lines within this region present elliptic topology referred to as O-type lines. In contrast, for $D_X < 0$ all three distinct eigenvalues are real, and hyperbolic magnetic field lines which are referred to as X-type lines can be found. More detailed classifications can be found in Perry & Chong (1987) and Chong et al. (1990). Consequently, the topological features of magnetic-field lines can be determined by the joint probability distributions (PDF) of $P(R_X, Q_X)$.

Additional information can be inferred by decomposing the magnetic field gradient tensor into its symmetric part $\mathbf{K} = \frac{1}{2}(\mathbf{X} + \mathbf{X}^T)$ referred as strain rate tensor and antisymmetric part $\mathbf{J} = \frac{1}{2}(\mathbf{X} - \mathbf{X}^T)$ associated with current density $\boldsymbol{j}$ ($\nabla \times \mathbf{B} = \boldsymbol{j}$) by $j_k = 2\epsilon_{ijk} J_{ij}$, here $\epsilon_{ijk}$ is the Levi-Civita symbol. Note the trace-less features of $\mathbf{K}$ and $\mathbf{J}$, two invariants can be defined as

$$Q_K = -\frac{1}{2}\text{Tr}(\mathbf{K}^2) \qquad (6)$$

and

$$Q_J = -\frac{1}{2}\text{Tr}(\mathbf{J}^2) = \frac{1}{4}\boldsymbol{j}^2, \qquad (7)$$

Subsequently, the joint PDF between $Q_K$ and $Q_J$ can be constructed to represent the spatial correlation between the current-associated energy and that related to the magnetic-field straining motions. Note that $Q_K$ is negative definitely, the joint PDF of $P(Q_J, -Q_K)$ is used instead.

Following the similar pattern, the geometric invariants of the velocity gradient tensor as well as its symmetric part and asymmetric part can be constructed. For incompressible plasma where $\nabla \cdot \mathbf{V} = \mathbf{0}$ (it is the same as the non-divergence form of magnetic field), these geometric invariants associated with velocity gradient tensor follow the same mathematical form as those of magnetic field. However, the compressible features make topology classification cannot be determined by the simple plane area division but a complicated spatial area division (e.g., Perry & Chong, 1987; Chong et al., 1990) which exceeds the study of this letter. The relation of velocity-field rotations and straining motions can be extracted from two geometrical invariants relying on the decomposition of the velocity-field gradient tensor $\nabla \mathbf{V}^\alpha$, here the superscript can be $i$ (ion) or $e$ (electron), representing plasma particles of type $\alpha$. The intrinsic decomposition of velocity-field gradient tensor is $\nabla \mathbf{V}^\alpha = \mathbf{S}^\alpha + \mathbf{\Omega}^\alpha = \frac{1}{3}(\nabla \cdot \mathbf{V}^\alpha)\mathbf{I} + \mathbf{D}^\alpha + \mathbf{\Omega}^\alpha$, where $\mathbf{S}^\alpha = \frac{1}{2}[\nabla \mathbf{V}^\alpha + (\nabla \mathbf{V}^\alpha)^T]$ and $\mathbf{\Omega}^\alpha = \frac{1}{2}[\nabla \mathbf{V}^\alpha - (\nabla \mathbf{V}^\alpha)^T]$ with $\omega_k{}^\alpha = 2\epsilon_{ijk}\mathbf{\Omega}_{ij}{}^\alpha$. Here, $\mathbf{S}^\alpha$ and $\mathbf{\Omega}^\alpha$ are the strain-rate and rotation-rate tensors of species $\alpha$, respectively; $\boldsymbol{\omega}^\alpha$ and $\mathbf{D}^\alpha$ are the vorticity ($\nabla \times \mathbf{V}^\alpha = \boldsymbol{\omega}^\alpha$) and the traceless strain-rate tensor. Note the traceless characteristics of $\mathbf{D}^\alpha$ and $\mathbf{\Omega}^\alpha$, the second geometrical invariants of them can be constructed as

$$Q_D{}^\alpha = -\frac{1}{2}\text{Tr}\,[(\mathbf{D}^\alpha)^2] \tag{8}$$

and

$$Q_\Omega{}^\alpha = -\frac{1}{2}\text{Tr}\,[(\mathbf{\Omega}^\alpha)^2] = \frac{1}{4}(\boldsymbol{\omega}^\alpha)^2. \tag{9}$$

As a consequence, the joint PDF of $P(Q_\Omega{}^\alpha, -Q_D{}^\alpha)$ can be established to represent the relation between the straining and rotational part of velocity gradient tensor.

## 3. Data Description and Results

To cover a large enough statistical sample in the turbulent reconnection outflow, we focus on a nearly 90-minute MMS burst-mode interval from 04:01 to 05:28 UT on May 28, 2017. This reconnection event has been investigated in previous studies (e.g., Zhou et al., 2021; Huang et al.,2022; Li et al., 2022). We use 128 Hz magnetic field data from the Flux Gate Magnetometer (FGM) instrument (Russell et al., 2016), 8192 Hz three-dimensional electric field data from the Electric Double Probes (Ergun et al., 2016; Lindqvist et al., 2016), 33 Hz electron data and 8 Hz ion data from the Fast Plasma Investigation instrument (Pollock et al., 2016) onboard MMS for this study.

A vital point of the topology classification lies in the construction of gradient tensor, and the nearly regular tetrahedron formed by four MMS spacecraft allows us to employ the curlometer technique (e.g., Dunlop et al. 1988) to build these gradient tensors. The separation among four MMS spacecrafts is about 60 km, i.e., ~0.2 ion-inertial lengths or ~6 electron-inertial lengths (ion inertial length of 354 km and electron inertial length of 9 km based on the background parameters: $B$ = 8.6 nT, $N_i = N_e$ = 0.3 cm$^{-3}$), which indicates that geometrical topology is at sub-ion scales (i.e., kinetic scales). The small elongation ($E \sim 0.12$) and planarity ($P \sim 0.11$) parameter values of the MMS tetrahedron configuration imply that the MMS forms very regular tetrahedron (Robert et al., 1998), thus guaranteeing the reliability of the results.

Figure 1 displays the time series of all normalized geometrical invariants and the parameter $\boldsymbol{j} \cdot \boldsymbol{E}'(\boldsymbol{E}' = \boldsymbol{E} + \boldsymbol{V}_e \times \boldsymbol{B})$ which has been selected as a proxy of energy dissipation in many previous studies (e.g., Zenitani et al., 2012; Wan et al., 2015; Burch et al., 2016b; Vörös et al., 2017; Chasapis et al., 2018; Huang et al. 2018, 2019, 2022; Jiang et al., 2019, 2021; Xiong et al., 2022). We normalize these invariants by $nR_X = R_X/<|\boldsymbol{j}|^2>^{3/2}$, $nQ_X = Q_X/<|\boldsymbol{j}|^2>$, $nQ_J = Q_J/<|\boldsymbol{j}|^2>$, $nQ_K = Q_K/<|\boldsymbol{j}|^2>$, $nQ_\Omega^e = Q_\Omega^e/<|\boldsymbol{\omega}^e|^2>$, $nQ_D^e = Q_D^e/<|\boldsymbol{\omega}^e|^2>$, $nQ_\Omega^i = Q_\Omega^i/<|\boldsymbol{\omega}^i|^2>$, $nQ_D^i = Q_D^i/<|\boldsymbol{\omega}^i|^2>$, where ⟨…⟩ represents the average on the whole interval. Through the time

variations of these invariants, it suggests that the intense energy dissipation (Figure 1c) is approximately correlated with both $nQ_X$ and $nR_X$ (Figure 1a-1b), implying the close relationship between coherent structures and energy dissipation. Besides, the magnetic field straining (Figure 1e) expresses a synchronized trend as energy-associated current structures (Figure 1d). Moreover, both electron (Figure 1f-1g) and ion (Figure 1h-1i) velocity tensors show the tight spatial connection with their straining part and rotational part. Next, we will show these correlations and the relationship between energy dissipation and topology and coherent structures.

The joint PDFs of $P(nR_X, nQ_X)$, $P(nQ_J, -nQ_K)$, $P(nQ_\Omega^e, -nQ_D^e)$, $P(nQ_\Omega^i, -nQ_D^i)$ are shown in Figure 2. In Figure 2a, one can see a cigar-like shape in the $P(nR_X, nQ_X)$ distribution. The solid magenta line is the separatrix satisfying the condition $D_X = 0$. Summing the counts above and below this line, it indicates that the ratio between O-type topologies ($D_X > 0$) and X-type topologies ($D_X < 0$) is 66.16%:33.84%. This dominant distribution of O-type topologies is consistent with the results observed in the solar wind (e.g., Quattrociocchi et al., 2019; Hnat et al., 2021) and the Earth's magnetosheath (e.g., Zhang et al., 2023; Ji et al., 2023). Figure 2b displays the joint PDFs of $P(nQ_J, -nQ_K)$, the dominant distribution near the bisector lines is very similar to the MHD-scales results in Quattrociocchi et al. (2019) and Dallas & Alexakis (2013). The linear correlation coefficient is 0.84, which implies a strong spatial correlation between these two invariants. For the PDF $P(nQ_\Omega^e, -nQ_D^e)$ and $P(nQ_\Omega^i, -nQ_D^i)$ (Figure 2c and 2d), the correlations seem to be weaker which are analogous to the results observed at Earth's magnetosheath (e.g., Bandyopadhyay et al., 2020; Zhang et al., 2023), and the two correlation coefficients are 0.58 and 0.82, respectively.

To estimate the spatial correlation between energy dissipation and above geometric invariants, we compute the conditional averages $\boldsymbol{j} \cdot \boldsymbol{E}'$ with these quantities. The results are shown in Figure 3. The conditions are based on values of $nQ_J$, $-nQ_K$, $nQ_\Omega^e$,

$nQ_D^e$, $nQ_\Omega^i$ and $nQ_D^i$. For example, to compute $<\boldsymbol{j} \cdot \boldsymbol{E}'|nQ_J>$, we calculate the average $\boldsymbol{j} \cdot \boldsymbol{E}'$ including only values occurring at spatial positions where $nQ_J$ exceeds a selected threshold. As can be seen in Figure 3a, elevated levels of $\boldsymbol{j} \cdot \boldsymbol{E}'$ are found in regions with enhanced magnetic stress and enhanced current density. This result can also be observed from the good correlations of time series between $\boldsymbol{j} \cdot \boldsymbol{E}'$ and $nQ_J$ as well as $-nQ_K$ in Figure 1c-1e. However, the averages of $\boldsymbol{j} \cdot \boldsymbol{E}'$ conditioned on the other invariants remain fairly constant. This suggests that Ohmic dissipation is more likely to occur in structures with strong magnetic stress or current density. Considering the results that strong incompressible pressure-strain interactions are associated with enhanced vorticity or symmetric stress (e.g., Yang et al., 2017; Bandyopadhyay et al., 2020), we can infer that forms of energy dissipation occurring in different structures may be various. To associate the conditional averages with magnetic-field topology classification in Figure 2a, we divide all the data into two parts as O-type topologies and X-type topologies. The conditional averages of $\boldsymbol{j} \cdot \boldsymbol{E}'$ on different magnetic field topologies are shown in Figure 3b-3c. Here we only consider the averages conditioned on $nQ_J$ and $-nQ_K$, and the results illustrate that the growth trend are mostly contributed by O-point topologies, which indicates that the energy dissipation prefers to occur in the O-point topologies in the turbulent reconnection outflow.

## 4. Conclusions and Discussions

In this work, we apply a local streamline-topology classification methodology in the turbulent reconnection outflow of the Earth's magnetotail. The characterization of magnetic-field topological structures at kinetic scales is exhibited, and it is found that the proportions of O-type topologies and X-type topologies are 66.16% and 33.84%, respectively. In addition, strong spatial correlations between the straining and rotational part of the velocity gradient tensor as well as the magnetic-field gradient tensor is observed. We also investigate the correlations between energy dissipation and these topological structures. The results indicate that energy dissipation occurs at regions with intense high magnetic stress or current sheets and mostly contributed by O-point topologies.

The observed result that Ohmic dissipation tends to occur at regions with high magnetic stress or current density supports the viewpoint that Ohmic dissipation occurs in current sheets (e.g., Dallas & Alexakis, 2013). Fu et al. (2017) have found that energy dissipation occurs at O-lines but not X-lines during one reconnection event. Our discovery that O-type topologies contribute much to energy dissipation is consistent with the result of Fu et al. (2017), which may indicate that O-type magnetic-field topologies are more likely to be the potential sites for energy dissipation than X-type topologies in both reconnection and turbulence. In future, since there are much higher quality data in the magnetosheath than in the magnetotail from MMS mission, we plan to conduct a much more comprehensive analysis in terrestrial magnetosheath to study the field gradient tensor features at kinetic scales, particularly focusing on relating energy conversion with different geometrical topologies.

**Data Availability**

The MMS data are available at https://lasp.colorado.edu/mms/sdc/public/data/.

**Acknowledgment**

This work was supported by the National Natural Science Foundation of China (42074196, 41925018) and the National Youth Talent Support Program. S.Y.H. acknowledges the project supported by Special Fund of Hubei Luojia laboratory.

**Appendix Error Analysis**

When computing the geometrical invariants, errors arise from the estimation of the gradient tensor. The gradient tensor calculated by the curlometer technique is defined as $\partial_i V_j = \sum_\alpha k_{\alpha,i} V_{\alpha,j}$, where $\alpha = 1,2,3,4$ denotes the spacecraft and $k_{\alpha,i}$ is the component of the reciprocal vector defined as $\boldsymbol{k}_\alpha = \frac{\boldsymbol{r}_{\beta\gamma} \times \boldsymbol{r}_{\beta\lambda}}{\boldsymbol{r}_{\beta\alpha} \cdot (\boldsymbol{r}_{\beta\gamma} \times \boldsymbol{r}_{\beta\lambda})}$, here $\boldsymbol{r}_{\alpha\beta} = \boldsymbol{r}_\beta - \boldsymbol{r}_\alpha$ are the relative position vectors of the four spacecraft, where $(\alpha, \beta, \gamma, \lambda)$ must be a cyclic permutation of $(1,2,3,4)$ (e.g., Chanteur, 1998). The primary uncertainty sources for velocity field geometrical invariants come from the plasma moments and the spacecraft tetrahedron's shape (e.g., Roberts et al., 2023). In this study, due to the small planarity parameter $P$ and elongation parameter $E$

($P\sim0.11$, $E\sim0.12$), the uncertainty caused by the tetrahedron's shape is then expected to be small compared to the errors from the plasma moments (e.g., Roberts et al., 2023), and we only discuss the possible effects from the plasma moments' errors on calculating the velocity field geometrical invariants. Then, the error of $\partial_i V_j$ can be denoted by

$$\sigma_{\partial_i V_j} = \sqrt{\sum_\alpha [k_{\alpha,i}^2 (\sigma_{V_{\alpha,j}})^2]}, \quad (10)$$

where $\sigma_{\partial_i V_j}$ denotes the error of $\partial_i V_j$, and the statistical errors $\sigma_{V_{\alpha,j}}$ are from FPI level-2 moments (e.g., Gershman et al. 2015, Pollock et al., 2016).

Applying the uncertainty propagation, one can obtain the errors of $Q_D = -\frac{1}{2}Tr(\boldsymbol{D}^2) = -\frac{1}{2}\left[\sum_{(i,j,k)\in\{(1,2,3),(2,3,1),(3,1,2)\}} \left(\frac{2}{3}\partial_i V_i - \frac{1}{3}\partial_j V_j - \frac{1}{3}\partial_k V_k\right)^2 + \frac{1}{2}\sum_{(i,j)\in\{(1,2),(1,3),(2,3)\}} \left(\partial_i V_j + \partial_j V_i\right)^2\right]$ and $Q_\Omega = \frac{1}{4}(\boldsymbol{\omega}^\alpha)^2 = \frac{1}{4}\sum_{(i,j)\in\{(1,2),(1,3),(2,3)\}} \left(\partial_i V_j - \partial_j V_i\right)^2$. For ease of representation, we define three intermediate variables as: $e_{ijk} = \left(\frac{2}{3}\partial_i V_i - \frac{1}{3}\partial_j V_j - \frac{1}{3}\partial_k V_k\right)^2$, $ed_{ij} = \left(\partial_i V_j + \partial_j V_i\right)^2$, $ew_{ij} = \left(\partial_i V_j - \partial_j V_i\right)^2$, and the errors of them can be respectively denoted as

$$\sigma_{e_{ijk}} = |e_{ijk}| \sqrt{2[(\tfrac{2}{3}\sigma_{\partial_i V_i})^2 + (\tfrac{1}{3}\sigma_{\partial_j V_j})^2 + (\tfrac{1}{3}\sigma_{\partial_k V_k})^2]}, \quad (11)$$

$$\sigma_{ed_{ij}} = |ed_{ij}| \sqrt{2[(\sigma_{\partial_i V_i})^2 + (\sigma_{\partial_j V_j})^2]}, \quad (12)$$

$$\sigma_{ew_{ij}} = |ew_{ij}| \sqrt{2[(\sigma_{\partial_i V_i})^2 + (\sigma_{\partial_j V_j})^2]}. \quad (13)$$

Then, one can obtain the errors of $Q_D$ and $Q_\Omega$ as:

$$\sigma_{Q_D} = \frac{1}{2}\sqrt{\sum_{(i,j,k)\in\{(1,2,3),(2,3,1),(3,1,2)\}} \left(\sigma_{e_{ijk}}\right)^2 + \frac{1}{4}\sum_{(i,j)\in\{(1,2),(1,3),(2,3)\}} (\sigma_{ed_{ij}})^2}, \quad (14)$$

$$\sigma[Q_\Omega] = \frac{1}{4}\sqrt{\sum_{(i,j)\in\{(1,2),(1,3),(2,3)\}} (\sigma_{ew_{ij}})^2}. \quad (15)$$

We performed two analytical methods to estimate the effects of errors on our results. The first method we used was to replace the original time series of $Q_D$ and $Q_\Omega$ by $Q_D + \sigma_{Q_D}$ and $Q_\Omega + \sigma_{Q_\Omega}$, subsequent processing is then implemented; the second method was to perform a statistical Monte Carlo test on the data to provide an additional estimate of the error. We took the individual velocity series and their respective errors and compute 100 new time series. This is performed by adding a random Gaussian error with a mean of zero and a standard deviation equal to the absolute statistical error to the measured velocity components. Next, we can obtain 100 series of the geometrical invariants $Q_D$ and $Q_\Omega$. Then we used the mean series of the 100 series to carry out the subsequent analysis. The results of the joint PDF of $P(nQ_\Omega, -nQ_D)$ and the conditional averages of $j \cdot E'$ by applying the two methods are presented in Figure 4a-4d. One can see that the joint PDFs of $P(nQ_\Omega, -nQ_D)$ is similar to the ones in Figure 2c and 2d. The correlation coefficients between $nQ_\Omega^e$ and $-nQ_D^e$ for these two methods are 0.64 and 0.54 (Figure 4a and 4b), respectively; and the correlation coefficients between $nQ_\Omega^i$ and $-nQ_D^i$ are 0.82 and 0.73 (Figure 4c and 4d), respectively. These coefficients are close to the ones not including errors (0.58 for $(nQ_\Omega^e, -nQ_D^e)$ and 0.82 for $(nQ_\Omega^i, -nQ_D^i)$ in Figure 2c and 2d, respectively). In addition, compared with the result in in Figure 3a, the influences of the two treatments on the conditional averaged results from the errors (Figure 4e and 4f) also reveal that the error analysis do not affect our conclusions. Therefore, all these results indicate that the errors from the plasma moments and the spacecraft tetrahedron's shape could not affect our conclusions in the present study.

**Figure captions**

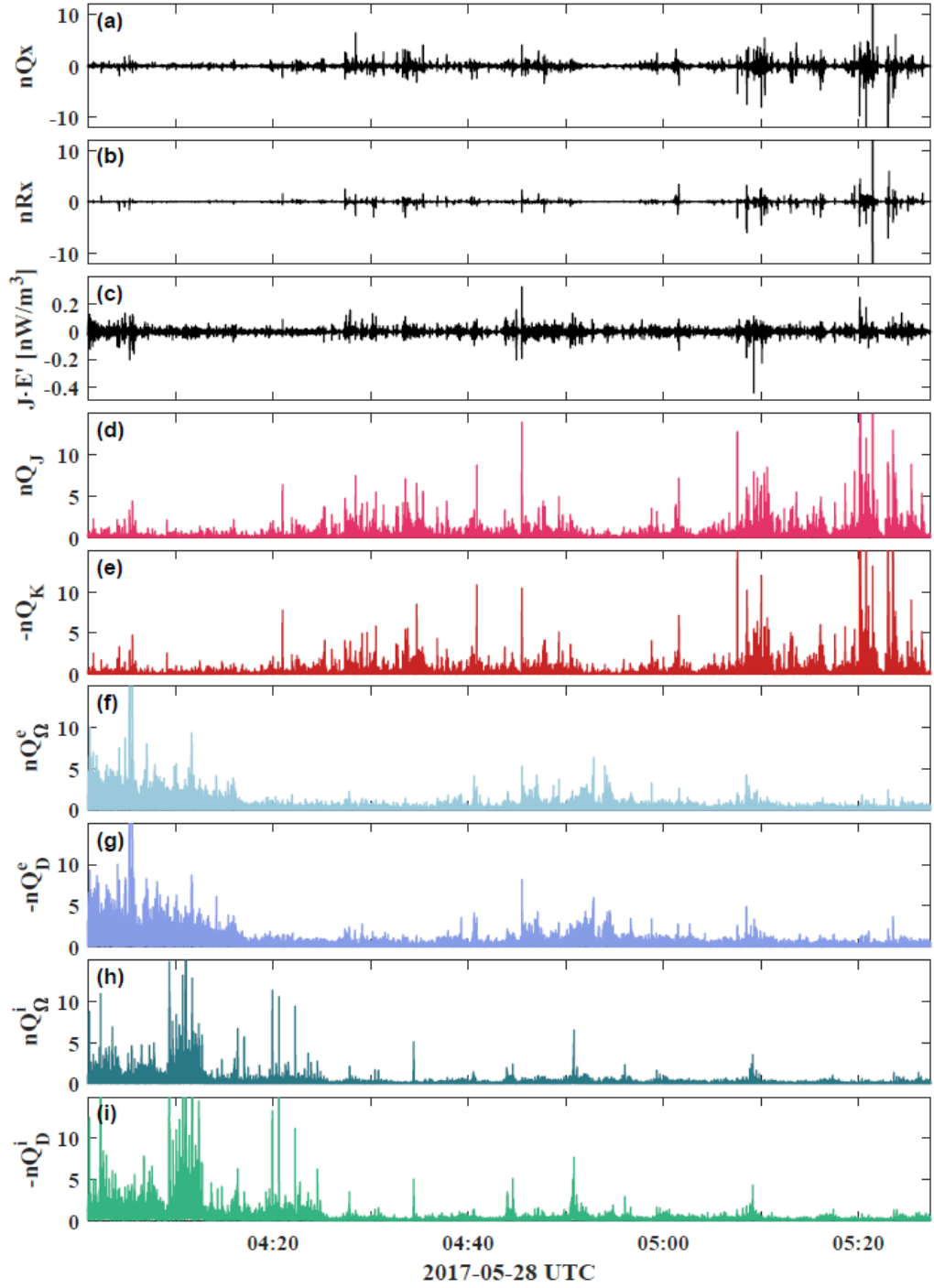

**Figure 1**. Time series of (a) $nR_X$, (b) $nQ_X$, (c) $\boldsymbol{j} \cdot \boldsymbol{E}'$, (d) $nQ_J$, (e) $-nQ_K$, (f) $nQ_\Omega^e$, (g) $-nQ_D^e$, (h) $nQ_\Omega^i$, (i) $-nQ_D^i$. The detailed definitions and normalizations can be found in the text.

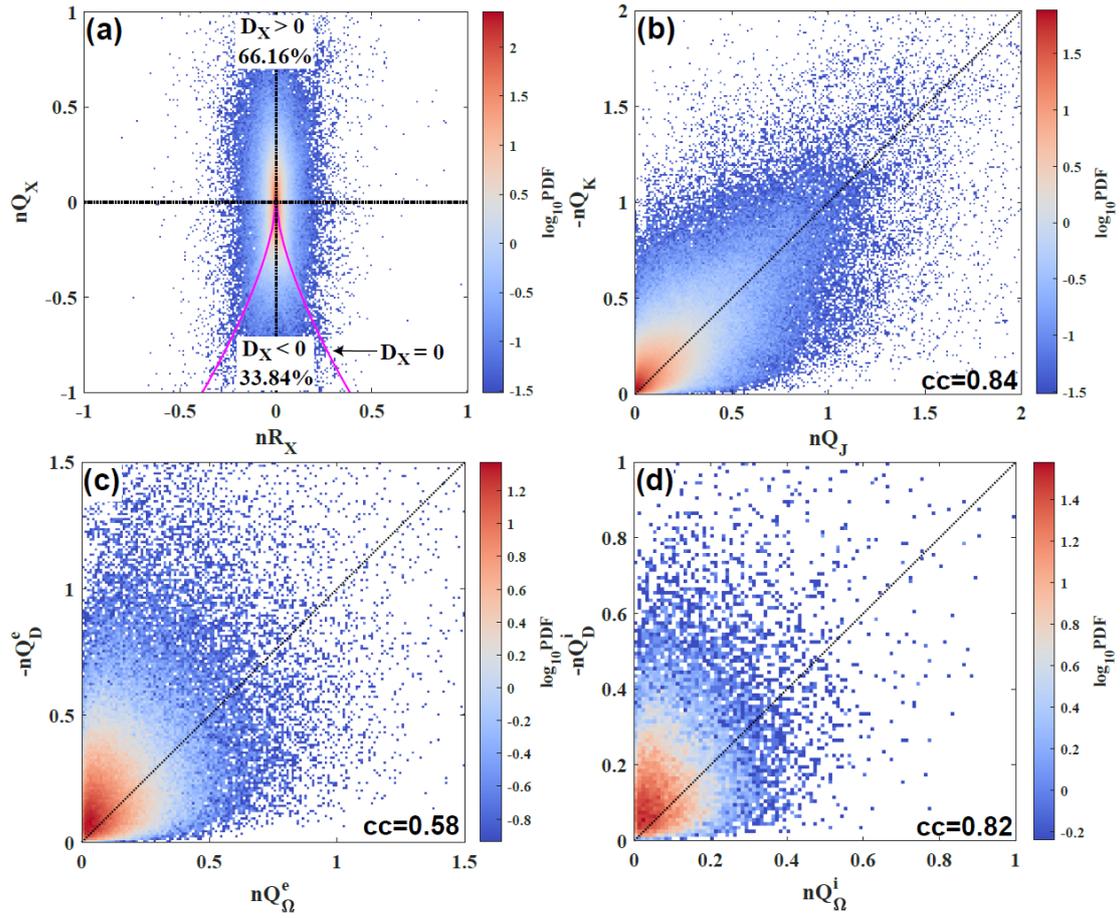

**Figure 2**. The joint PDF of (a) $P(nR_X, nQ_X)$, (b) $P(nQ_J, -nQ_K)$, (c) $P(nQ_\Omega^e, -nQ_D^e)$, (d) $P(nQ_\Omega^i, -nQ_D^i)$. The magenta line refers to the discriminant line $D_X = 0$.

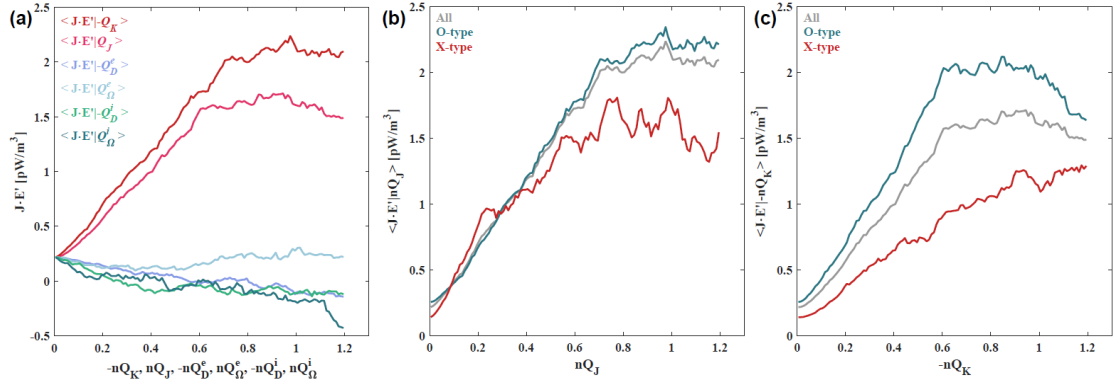

**Figure 3**. Conditional averages of $\mathbf{j}\cdot\mathbf{E}'$ on (a) six geometrical invariants, (b) $-nQ_K$ and (c) $nQ_J$ under three different conditions.

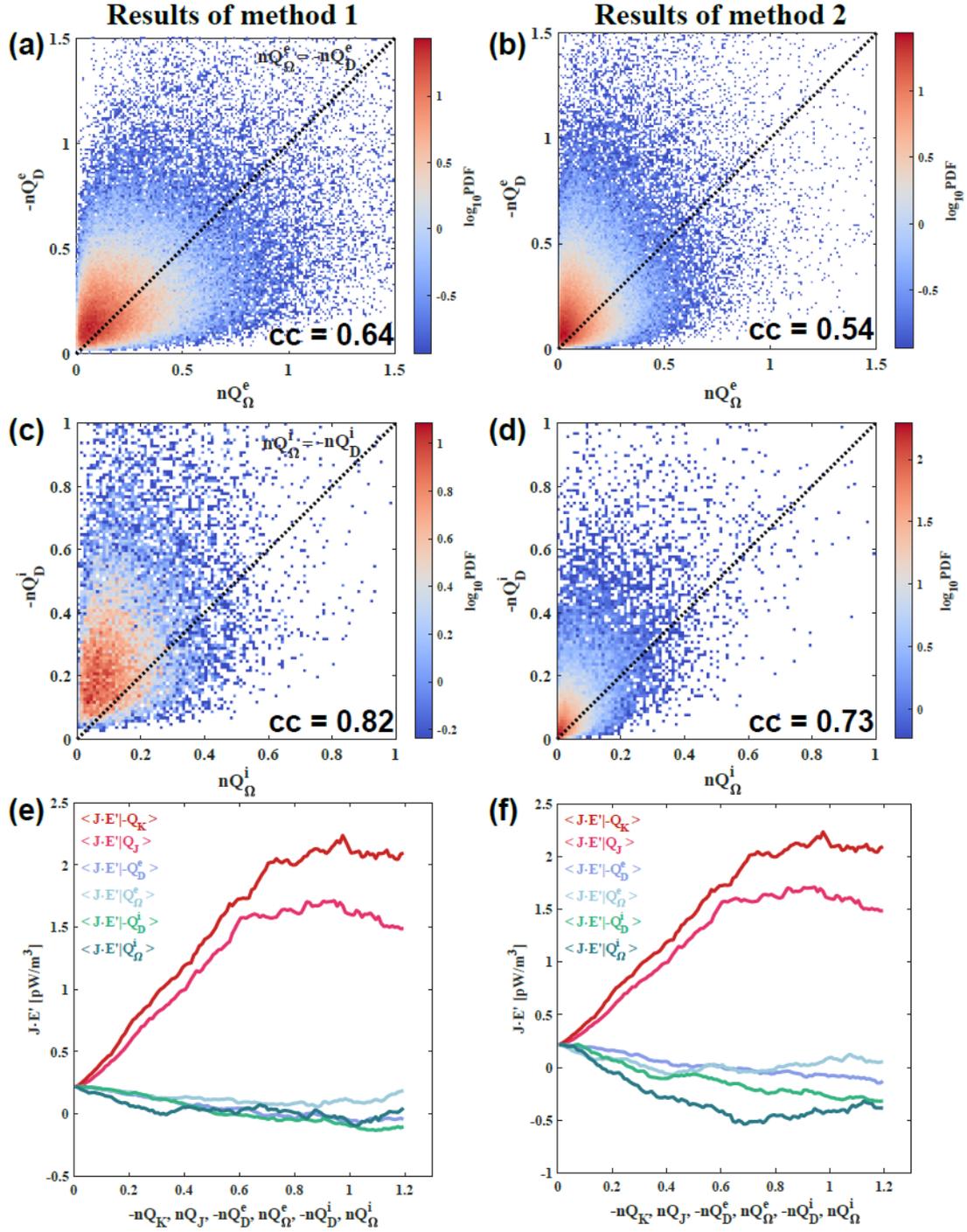

**Figure 4.** The first and the second rows represent the joint probability distribution (PDF) of $P(nQ_\Omega^e, -nQ_D^e)$, the joint PDF of $P(nQ_\Omega^i, -nQ_D^i)$ and the conditional averages of $\boldsymbol{j} \cdot \boldsymbol{E'}$ on six geometrical invariants. The left and right columns represent the results of method 1 and method 2, respectively.